\documentclass[preprint,epsfx,floatfix]{revtex4}
\usepackage{graphicx}
\usepackage{enumerate}
\usepackage{float}
\begin{document}
\title{Realization of Morphing Logic Gates in a Repressilator with
  Quorum Sensing Feedback} 
\author{Vidit Agarwal, Shivpal Singh Kang and Sudeshna Sinha}
\affiliation{Indian Institute of Science Education and Research
  (IISER) Mohali, Knowledge City, SAS Nagar, Sector 81, Manauli PO 140
  306, Punjab, India}

\begin{abstract}
  We demonstrate how a genetic ring oscillator network with quorum
  sensing feedback can operate as a robust logic gate. Specifically we
  show how a range of logic functions, namely AND/NAND, OR/NOR and
  XOR/XNOR, can be realized by the system, thus yielding a versatile
  unit that can morph between different logic operations. We further
  demonstrate the capacity of this system to yield complementary logic
  operations in parallel. Our results then indicate the computing
  potential of this biological system, and may lead to bio-inspired
  computing devices.

\end{abstract}

\maketitle
\section{Introduction}

The operation of any computing device is necessarily a physical
process, and this fundamentally determines the possibilities and
limitations of the computing machine. A common thread in the history
of computers is the exploitation and manipulation of different natural
phenomena to obtain newer forms of computing paradigms
\cite{computing}. For instance, chaos computing \cite{chaogates},
neurobiologically inspired computing, quantum computing\cite{quantum},
and DNA computing\cite{dna} all aim to utilize, at the basic level,
some of the computational capabilities inherent in natural systems. In
particular, larger understanding of biological systems has triggered
the interesting question: what new directions do bio-systems offer for
understanding and implementing computations?

The broad idea then, is to create machines that benefit from natural
phenomena and utilize patterns inherent in systems to encode inputs
and subsequently obtain a desired output. Different physical
principles can yield logic outputs, for logic gates such as AND, OR,
XOR, NOR and NAND which form the basis of the universal
general-purpose computation. It is particularly interesting if systems
of biological relevance can also yield logic outputs consistent with
the truth tables of different logic functions \cite{benenson, ando,
  gene2}.

In this work we will explore the computational possibilities arising
from the dynamics of a genetic network with quorum sensing
feedback. Quorum sensing allows cells to sense their own density and
to communicate with each other, giving the cells the ability to behave
as a population instead of individually. The autoinduction of
luminescence in the symbiotic marine bacterium \emph{Vibrio fischeri}
is perhaps the best characterized example of quorum sensing. Some
other examples include sporulation and fruiting body formation by
\emph{Myxococcus Xanthus} and antibiotic production by several species
of \emph{Streptomyces}. The way quorum sensing is accomplished in
cells is through diffusive exchange of auto-inducer molecules which
participate in intercellular coupling and self feedback. At low
cell density, the auto inducer is synthesized at basal levels and
diffuses into the extracellular space and gets diluted. But as the
cell density increases, the intracellular autoinducer concentration
increases until it reaches a threshold beyond which it is produced
auto-catalytically. This results in a dramatic increase of product
concentrations, thus creating bi-stability between a high steady
state and a low steady state \cite{dockery, james}.

Specifically, in this work we focus on an artificial genetic network
known as the repressilator, with feedback of the kind that quorum
sensing may provide to an isolated repressilator. A repressilator
consists of a ring of three genes, products of which inhibit each
other in cyclic order. The quorum sensing feedback stimulates the
activity of a chosen gene providing competition between the inbitory
and the stimulatory activities localized in that gene. In this work we
attempt to use the varying concentrations of proteins in the
repressilator, arising due to quorum sensing feedback, to realize
different logic gates.

In the sections below we will first discuss the model of the
repressilator genetic network with quorum sensing feedback, and
present the change in dynamics of repressilator proteins with
changing parameters of the quorum sensing feedback. We will then go on
to use this dynamics to implement a range of logic functions, by
appropriate input-output associations.

\section{Repressilator Genetic Network with Quorum sensing Feedback}

We consider a basic genetic ring oscillator network in which three
genes inhibit each other in unidirectional manner, and an additional
quorum sensing feedback loop stimulates the activity of a chosen gene,
providing competition between inhibitory and stimulatory activities
localized in that gene. This system is known to yield behaviour
ranging from limit cycles, to high and low stable steady states
\cite{Hellen}. 

In the model, 3 genes in a cyclic loop forms mRNA ($a,b,c$) and
proteins ($A,B,C$). The three genes inhibit each other, where one gene
is inhibited by the preceding gene. Quorum sensing feedback acts between
gene $B$ and $C$, where $B$ stimulates the production of $C$ by quorum
sensing along with an inhibitory action. The model can be reduced
to fast mRNA kinetics, where $a,b,c$ is assumed to be in a steady state
with their respective proteins ($A,B,C$), leading to $\frac{da}{dt} =
\frac{db}{dt} =\frac{dc}{dt} \approx 0$.

The equations describing the dynamics of this system then are:

\begin{eqnarray}
\frac{dA}{dt} &=& \beta_1 .( -A + \frac{\alpha}{1+C^n}) \\
\frac{dB}{dt} &=& \beta_2 .( -B + \frac{\alpha}{1+A^n})  \\ 
\frac{dC}{dt} &=& \beta_3 .( -C + \frac{\alpha}{1+B^n} + \frac{\kappa S}{1+S}) \\ 
\frac{dS}{dt} &=&  k_{S0}.S - k_{S1}.B - \eta .(S - S_{ext}) 
\end{eqnarray}

Here $S$ denotes the concentration of the auto-inducer AI.

The relevant parameters in this system are the ratio of protein decay
rate to mRNA decay rate $\beta_i$, maximum transcription rate in the
absence of an inhibitor $\alpha$, Hill co-efficient for inhibition
$n$, diffusion co-efficient depending on the permeability of the
membrane to auto-inducer molecule $\eta$, the rate of auto-inducer
decay to mRNA decay rate $k_{SO}$, rate of auto-inducer production
$k_{S1}$, and most importantly, the co-efficient for the production of
$C$ protein due to auto-inducer $\kappa$ and the concentration of
auto-inducer in external medium $S_{ext}$ responsible for the {\em
  quorum sensing feedback}.

In our simulations we fix the parameters at realistic values
\cite{Hellen}: $k_{SO} = 1.0$, $k_{S1} = 0.025$; $\eta = 1.0$; $n = 3$,
and $\beta_1 = \beta_2 = \beta_3 = \beta = 0.1$ (relevant to slow
protein kinetics).
We vary the parameters that determine the strength of quorum sensing,
namely $S_{ext}$ and $\kappa$.

\begin{figure}[h]
\begin{center}
\includegraphics[height = 2.5in,width = 3.2in]{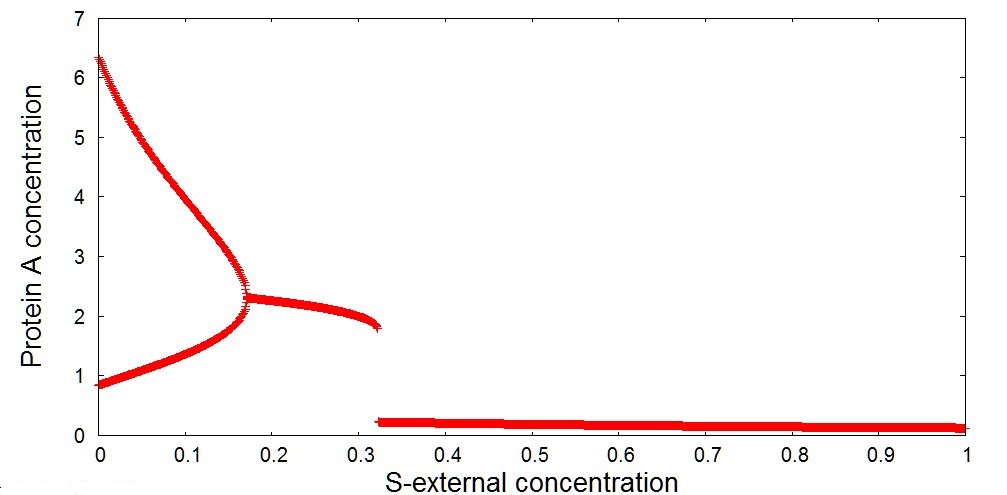}
\includegraphics[height = 2.5in,width = 3.2in]{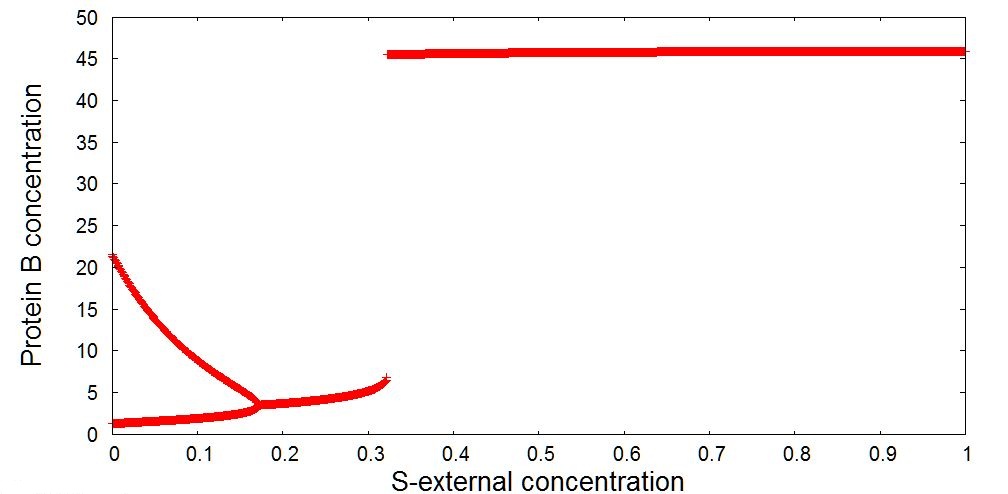}
\end{center}
\caption {Bifurcation diagram displaying the extremal values of the
  concentration of Protein $A$ (left) and Protein $B$ (right) vs the
  auto-inducer concentration in the external medium $S_{ext}$.  The
  values of other parameters are: $\beta = 0.1$, $\alpha = 46$ and
  $\kappa = 14$. For every set of input parameters, the
  concentrations of proteins are initialised with
  $A_{int}=B_{int}=C_{int}=0.0$.\\}
\label{bifSext}
\end{figure}

\begin{figure}[htb]
\begin{center}
\includegraphics[height = 2.5in,width = 3.2in]{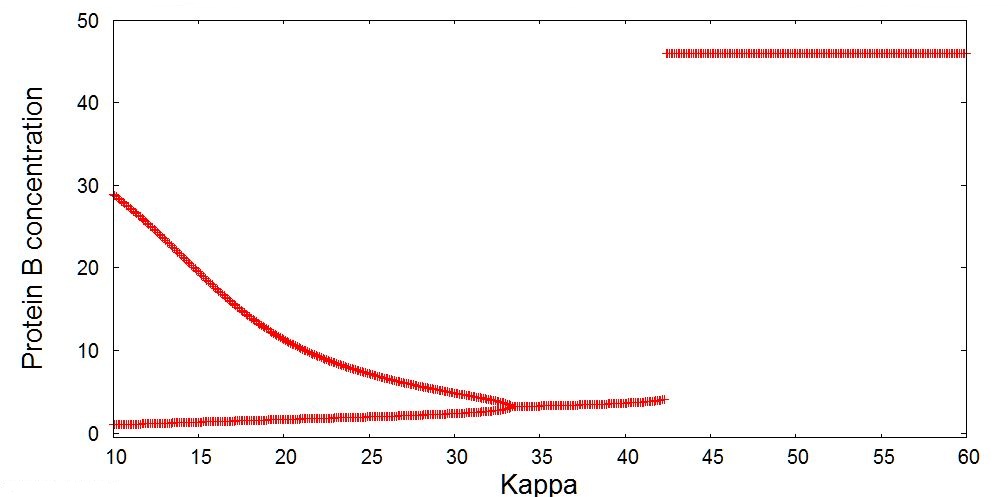}
\includegraphics[height = 2.5in,width = 3.2in]{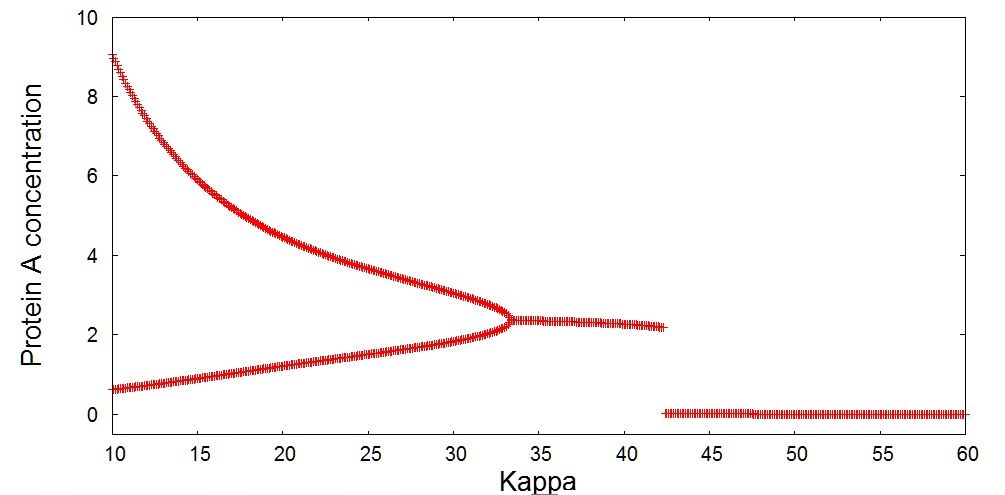}
\end{center}
\caption {Bifurcation diagram displaying the extremal values of the
  concentration of Protein $A$ (left) and Protein $B$ (right) vs
  parameter $\kappa$.  The values of other parameters are: $\beta =
  0.1$, $\alpha = 46$ and $S_{ext} = 0.0$. Here the variation of the
  strength of the quorum sensing feedback solely depends on $\kappa$.
  For every set of input parameters, the concentrations of proteins
  are initialised with $A_{int}=B_{int}=C_{int}=0.0$. }
\label{bifkappa}
\end{figure}

On taking different values of parameters $S_{ext}$ and $\kappa$, the
strength of quorum sensing changes, and hence the qualitative nature
of the dynamics changes. This is clearly evident from the bifurcation
analysis of the concentrations of protein $A$ and $B$ with respect to
parameters $S_{ext}$ and $\kappa$ displayed in
Figs. \ref{bifSext}-\ref{bifkappa}, which shows the change in dynamics
from limt cycles to stable steady states as the external concentration of
auto-inducer increases. There is also a sudden switch in the value of
the steady state after a critical value of $S_{ext}$ and $\kappa$.

\section{Realization of Logic Gates}

Now, in order to utilize this system to realize a logic gate we need
to propose an appropriate {\em input-output mapping}, namely, we must
first suggest how inputs and outputs are encoded. The task at hand
then, is to ensure that this input and output encoding {\em
  consistently yields the correct input-output association
  corresponding to different logic functions}, as displayed in the
truth table (Table I).

Specifically, here we will implement a series of logic gates, such as
AND/NAND, OR/NOR and XOR/XNOR, using the change in dynamics of protein
$A$ and $B$. Further we will obtain logic operations in parallel by
simultaneous measurement of the concentrations of these two proteins.\\

\bigskip

\textbf{Encoding Logic Inputs:}\\

Consider the inputs to be encoded by $S_{ext}$, namely the
concentration of auto-inducer in external medium, which is responsible
for the quorum sensing feedback.

We propose that the two inputs $I_1 + I_2 = S_{ext}$
where $I_1$ and $I_2$ encode the two logic inputs as follows:

\begin{enumerate}[(i)]
\item When logic input is $0$, $I_{1,2} = 0.0$;

\item When logic input is $1$, $I_{1,2} = 0.2$;
\end{enumerate}

So the pair of inputs lead to the following three distinct conditions:

\begin{enumerate}[(i)]

\item $S_{ext} = 0.0$  \ \ \ \ {\rm for  \ input \ set} \ \ \ $(I_1,I_2) \equiv (0,0)$

\item $S_{ext} = 0.2$   \ \ \ \ {\rm for  \ input \ set} \ \ \ $(I_1,I_2) \equiv (0,1)/(1,0)$

\item $S_{ext} = 0.4$   \ \ \ \ {\rm for  \ input \ set} \ \ \ $(I_1,I_2) \equiv (1,1)$\\
\end{enumerate}

\bigskip

\textbf{Encoding Logic Output:}\\


We consider the {\em maximum protein concentration as an indicator of
  the output of the system}. Specifically we consider the maximum
concentrations of proteins $A$ and $B$, denoted by $A_{max}$ and
$B_{max}$ respectively. This gives us the ability to obtain two
outputs from the system in parallel.

Now a prescribed output determination threshold, $A^*$ and $B^*$,
leads to the mapping of the concentration to a $0/1$ logic output, as
folllows: 

For protein $A$:

\begin{enumerate}[(i)]
\item If $A_{max}  < A^*$, then the Logic Output is $0$

\item If $A_{max} \ge A^*$, then the Logic Output is $1$
\end{enumerate}

And for protein $B$:

\begin{enumerate}[(i)]
\item If $B_{max} < B^*$, then the Logic Output is $0$

\item If $B_{max} \ge B^*$, then the Logic Output is $1$
\end{enumerate}

One can obtain the {\em complementary gates} by exchanging the output
determination criterion. Namely:

For protein $A$:

\begin{enumerate}[(i)]
\item If $A_{max}  \ge A^*$, then the Logic Output is $0$

\item If $A_{max} < A^*$, then the Logic Output is $1$
\end{enumerate}

And for protein $B$:

\begin{enumerate}[(i)]
\item If $B_{max} \ge B^*$, then the Logic Output is $0$

\item If $B_{max} < B^*$, then the Logic Output is $1$\\
\end{enumerate}

\bigskip

{\bf Controlling the nature of the Logic Function}:\\


We can control the type of logic operation obtained by simply changing
the output determination threshold. Namely, we can obtain AND/NAND,
OR/NOR or XOR/XNOR input-output mappings by considering different
$A^*$ and $B^*$.

\begin{figure}[htb]
\begin{center}
\includegraphics[height = 2.75in,width = 4in]{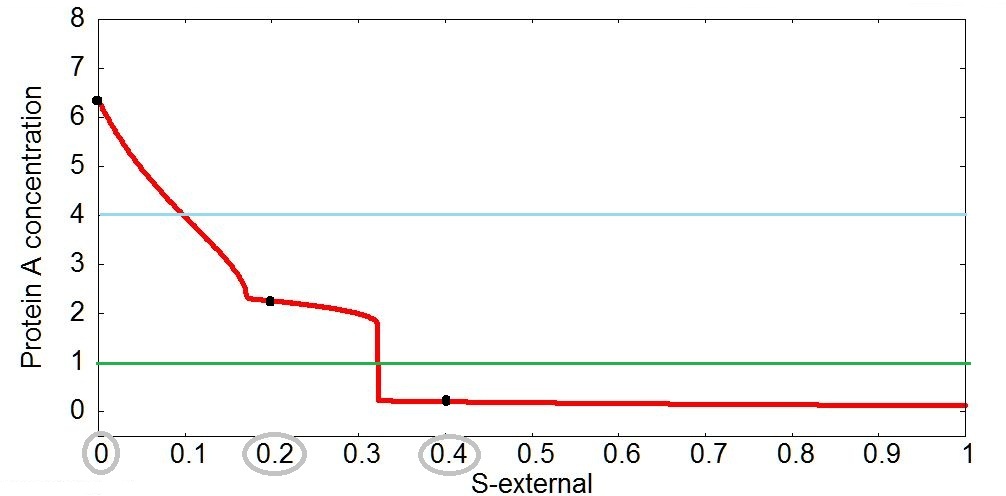}
\end{center}
\caption {Maximum concentration of protein $A$ vs. $S_{ext}$. The
  maximum is determined from a range of around 100 time steps. The
  fundamental NAND and NOR gates are indicated here, with the output
  determination level being $A^*=1$ (shown by green line) for NAND,
  and $A^*=4$ (shown by blue line) for NOR logic. The grey encircled
  values are the logic inputs, with the first circle at $S_{ext} = 0$
  denoting the input set $(0,0)$, the second circle at $S_{ext} = 0.2$
  denoting the input sets $(1,0)/(0,1)$, and the third circle at
  $S_{ext} = 0.4$ denoting the input set $(1,1)$ (see text for input
  encoding method). Other parameters are: $k_{SO} = 1.0$; $k_{S1} =
  0.025$; $\eta = 1.0$; $n = 3$, $\beta = 0.1$, $\alpha = 46$ and
  $\kappa = 14$.}
\label{NAND_A}
\end{figure}

For instance, for realizing the fundamental NAND Logic gate (see Truth
table I), we have the binary logic output determined by the threshold
level $A^* = 1$. Namely, if the maximum concentration of protein $A$,
$A_{max}$ is greater than $A^*$, then the Logic Output is $1$, and if
$A_{max} \le A^*$, then Logic Output is $0$. This is clearly evident
in Fig. \ref{NAND_A} and Table \ref{NAND_A_table}. 

In order to change the logic function from the fundamental NAND gate
to the funademental NOR gate, we simply have to change the prescribed
output determination threshold. Specifically, the system yields NOR
logic when the output determination threshold $A^*$ is $4$.

\begin{figure}[htb]
\center{\includegraphics[height = 2.75in,width = 4in]{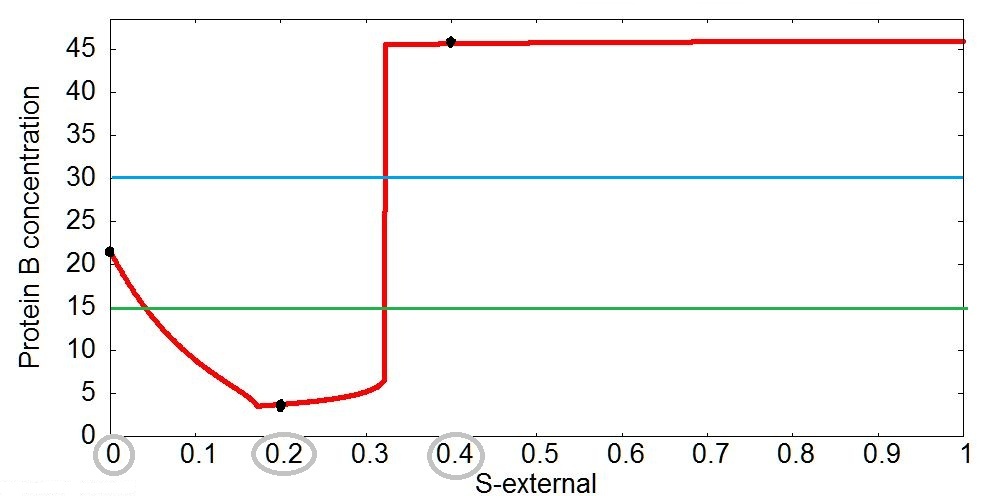}}
\caption {Maximum concentration of protein $B$ vs. $S_{ext}$. The
  maximum is determined from a range of around 100 time steps. The AND
  and XNOR gates are indicated here, using output determination
  threshold $B^* = 30$ and $B^* = 15$ respectively.  Again, the grey
  encircled values are the logic inputs, with the first circle at
  $S_{ext} = 0$ denoting the input set $(0,0)$, the second circle at
  $S_{ext} = 0.2$ denoting the input sets $(1,0)/(0,1)$, and the third
  circle at $S_{ext} = 0.4$ denoting the input set $(1,1)$ (same input
  encoding method as in Fig. \ref{NAND_A}). Other parameters are:
  $k_{SO} = 1.0$; $k_{S1} = 0.025$; $\eta = 1.0$; $n = 3$, $\beta =
  0.1$, $\alpha = 46$ and $\kappa = 14$.}
\label{AND_B}
\end{figure}

Similarly, using different output determination thresholds $B^*$ for
protein concentration $B$, we can obtain AND and XNOR logic
operations. This is clearly evident through Fig. \ref{AND_B} and Table
\ref{AND_B_table}. Again note that the complementary logic functions
NAND and XOR are implemented by a simple toggle of the logic output
determination condition.

\begin{figure}[htb]
\begin{center}
\includegraphics[height = 2.5in,width = 3.2in]{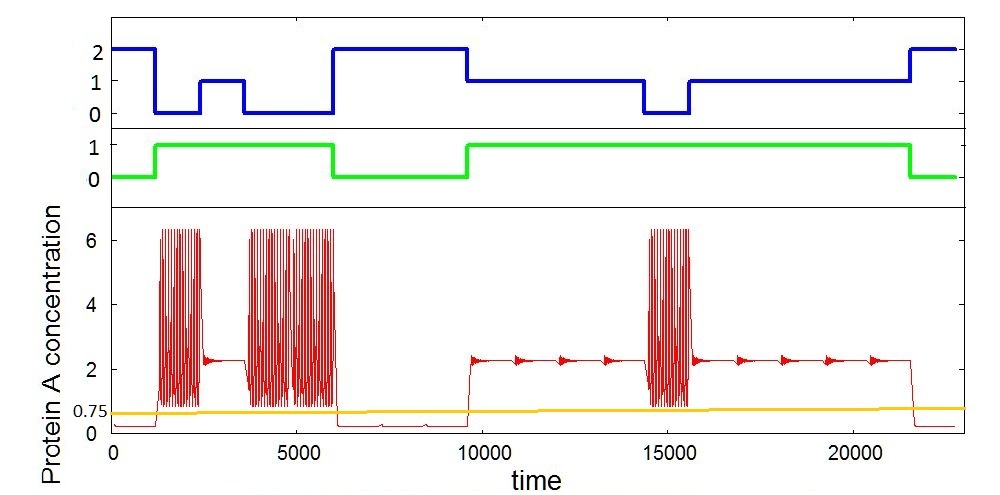}
\includegraphics[height = 2.5in,width = 3.2in]{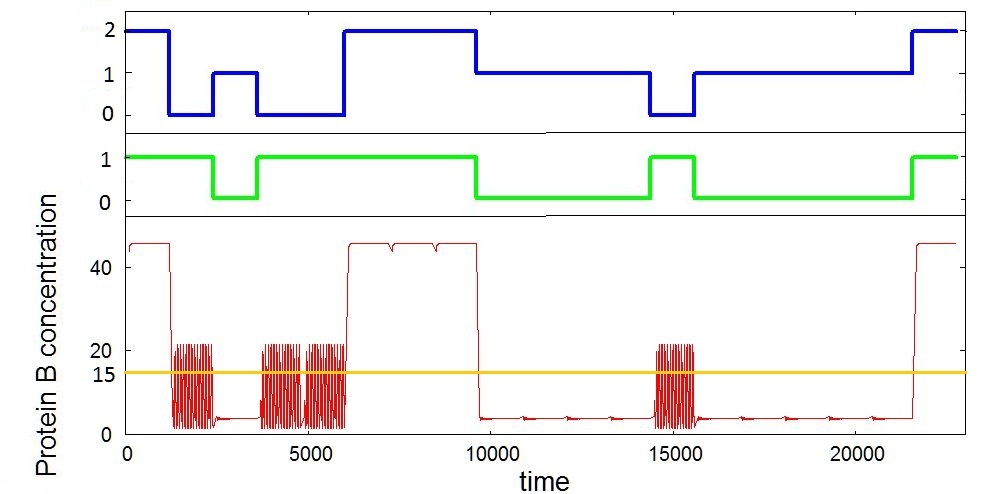}
\end{center}
\caption { Time evolution of the concentrations of Protein $A$ (left)
  and Protein $B$ (right). Here the output determination threshold
  (shown as a yellow line), which serves as logic selector, is set at
  $A^* = 0.75$ and $B^* = 15$. The timing sequence in blue (top) shows
  a random stream of binary logic input values $I_1 + I_2$. The
  sequence in green (middle) displays the binary logic output desired
  for the implementation of NAND (left) and XNOR (right) gates. Note
  that every time a new input set is taken, the transient dynamics
  ($\approx 100$), indicative of the latency, is not displayed.}
\label{timeseries}
\end{figure}

So we can obtain any combination of AND/NAND/OR/NOR in parallel with
AND/NAND/XOR/XNOR logic through the two protein concentrations for the
same pair of inputs. In Fig. \ref{timeseries}, the time series plots
for Protein $A$ and $B$ exhibits the simultaneous realization of NAND
and XNOR gates respectively, for a representative random stream of
input values. Clearly, this same system can yield XOR and AND in
parallel by the alternate output association, namely a logic output
$1$ is obtained when the proteins $A$ and $B$ are below output
threshold $A^*$ and $B^*$ respectively, and $0$ otherwise. In fact,
the combination of AND and XOR in parallel is particularly useful, as
it forms the basis of the ubiquitous {\em bit-by-bit addition}.

Alternately, the logic operation may also be changed by a constant
shift in input encoding, namely $I_1 + I_2 = S_{ext} + S_{logic}$,
where different values of $S_{logic}$ yield different logic
functions. For instance, using protein $B$ for output determination,
and keeping level $B^*$ fixed at $15$, $S_{logic} = 0$ yields XNOR,
and $S_{logic} = 0.1$ yields NAND. Similarly, using protein $A$ for
output determination, with $A^*=1$, $S_{logic} = 0$ yields NAND, and
$S_{logic} = 0.2$ yields NOR.\\

\bigskip

{\bf Using activation rate parameter $\kappa$ to encode Logic Inputs}:\\

\begin{figure}[htb]
\begin{center}
\includegraphics[height = 2.75in,width = 4in]{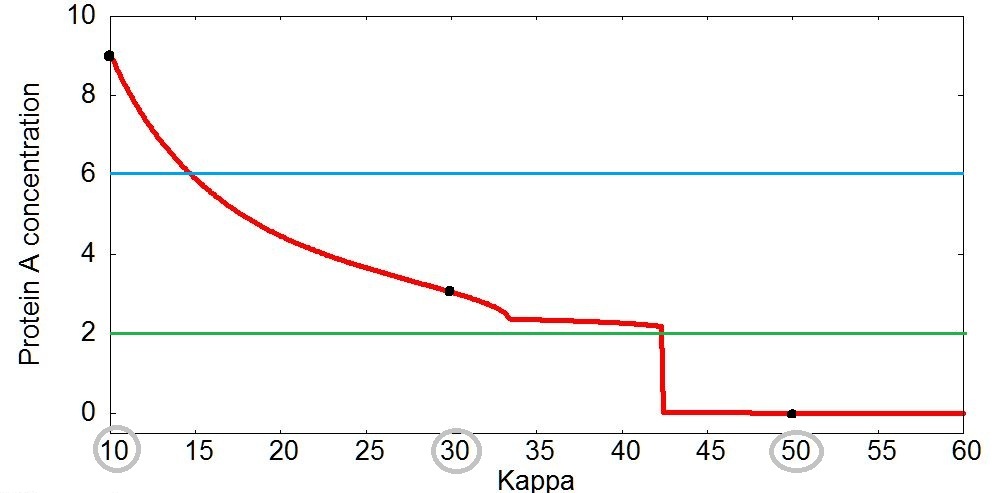}
\end{center}
\caption {Maximum concentration of protein $A$ vs. $\kappa$. The
  maximum is determined from a range of around 100 time steps. The
  fundamental NAND and NOR gates are indicated here, using output
  determination threshold $A^* = 2$ and $A^* = 6$ respectively.  The
  grey encircled values are the logic inputs, with the first circle at
  $\kappa = 10$ denoting the input set $(0,0)$, the second circle at
  $\kappa = 30$ denoting the input sets $(1,0)/(0,1)$, and the third
  circle at $\kappa = 50$ denoting the input set $(1,1)$ (see text for
  input encoding method). Other parameters are: $k_{SO} = 1.0$;
  $k_{S1} = 0.025$; $\eta = 1.0$; $n = 3$, $\beta = 0.1$, $\alpha =
  46$ and $S_{ext} = 0$.}
\label{kappa_A}
\end{figure}
\begin{figure}[htb]
\begin{center}
\includegraphics[height = 2.75in,width = 4in]{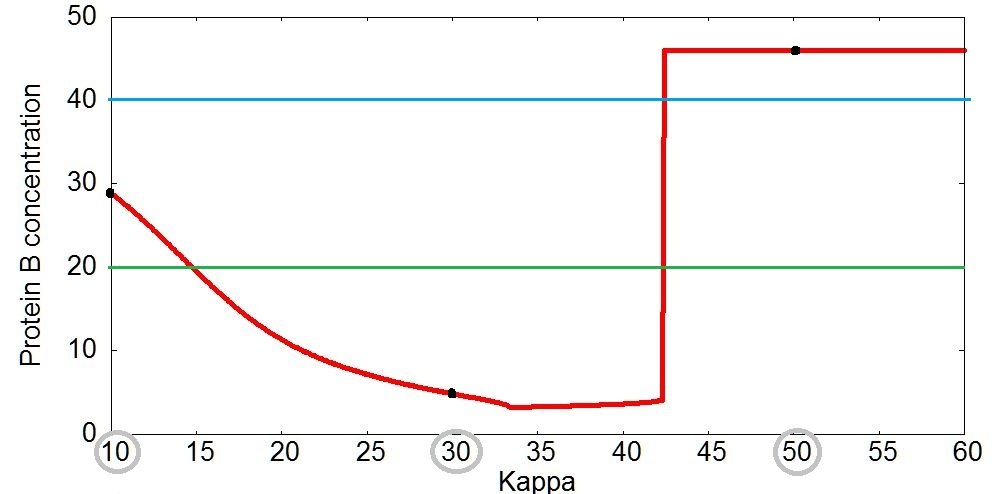}
\end{center}
\caption {Maximum concentration of protein $B$ vs. $\kappa$. The
    maximum is determined from a range of around 100 time steps. The
    AND and XNOR gates are indicated here, using output determination
    threshold $B^* = 40$ and $B^* = 20$ respectively.  The grey encircled
  values are the logic inputs, with the first circle at $\kappa = 10$
  denoting the input set $(0,0)$, the second circle at $\kappa = 30$
  denoting the input sets $(1,0)/(0,1)$, and the third circle at
  $\kappa = 50$ denoting the input set $(1,1)$ (see text for input
  encoding method). Other parameters are: $k_{SO} = 1.0$; $k_{S1} =
  0.025$; $\eta = 1.0$; $n = 3$, $\beta = 0.1$, $\alpha = 46$ and
  $S_{ext} = 0$.}
\label{kappa_B}
\end{figure}

We now consider an alternate input encoding, with the logic inputs
determining $\kappa$. Here $S_{ext}$ is held fixed at $0.0$, and
varying $\kappa$ yields different protein concentrations, which leads
to different outputs (see Figs. \ref{kappa_A}-\ref{kappa_B}).

Specifically, let $\kappa$ be determined by the logic inputs $I_1$ and
$I_2$ as follows: $$\kappa = I_1 + I_2 + \kappa_{base}$$ with $I_{1,2} =
0$ when logic input is $0$, and $I_{1,2} = 20$ when logic input is
$1$.

Taking a base value of $\kappa_{base}  = 10$, we have:

\begin{enumerate}[(i)]
\item $\kappa = 10$  when input set  ($I_1, I_2$) is $(0,0)$ 

\item $\kappa = 30$  when input set  ($I_1, I_2$) is $(0,1)/(1,0)$ 

\item $\kappa = 50$  when input set  ($I_1, I_2$) is $(1,1)$ 
\end{enumerate}

The binary Logic Output is obtained from the maximum concentration of
proteins $A$ and $B$, as before. Namely, we determine the logic output
through a threshold: if $A_{max} (B_{max}) < A^* (B^*)$, then logic
output is $0$, and if $A_{max} (B_{max}) < A^* (B^*)$, then logic
output is $1$.

As before, the nature of the logic function obtained can be controlled
by the output determination threshold for the proteins $A$ and
$B$. When $A^* = 2$ we obtain NAND logic and for $A^*=6$ we obtain the
fundamental NOR operation. Similarly, when $B^* = 40$ we obtain AND
and for $B^*=20$ we obtain XNOR. Representative values are displayed
in Tables \ref{kappa_A_table} and \ref{kappa_B_table}.

Hence, XNOR/XOR/AND/NAND and NAND/AND/NOR/OR gates can be realized in
parallel for the same set of inputs. Namely, by setting the
appropriate output determination condition, the protein concentrations
can yield any of the logic functions XNOR/XOR/AND/NAND and
NAND/AND/NOR/OR independently and simultaneously.

\section{Conclusions}

In summary, we have demonstrated how a genetic ring oscillator network
with quorum sensing feedback can operate as a robust logic gate. We
use the concentration of auto-inducer in external medium $S_{ext}$,
and alternatively an activation rate parameter $\kappa$ (cf. Eqn. 1),
to encode logic inputs. Both these quantities regulate the strength of
the quorum sensing feedback. The concentrations of proteins $A$ and
$B$ determine the logic outputs. This input-output association yields
a range of logic functions, namely AND/NAND, OR/NOR and XOR/XNOR.  So
the system can act as a versatile unit that can morph between
different logic operations. We further demonstrated the capacity of
this system to yield different logic operations simultaneously through
the two proteins $A$ and $B$.


These observations may provide an understanding of the computational
capacity of systems with quorum sensing feedback. It also may have
potential relevance in the design of biological gates, with the
ability to flexibly reconfigure logic operations, and implement logic
operations in parallel. Further, since electronic analogs of such
systems have already been implemented \cite{Hellen}, our results may
readily lead to bio-inspired computing devices.\\

\newpage
\begin{table}
\begin{center}
\begin{tabular}{| c | c | c | c | c | c | c |}
 \hline Inputs $(I_1,I_2)$ &AND&OR&NAND&NOR&XOR&XNOR\\
\hline
(0,0) & 0  & 0  & 1 & 1  & 0  &  1  \\ \hline
(0,1)/(1,0) & 0  & 1  & 1  & 0  & 1 & 0  \\ \hline
(1,1)  & 1  & 1  & 0  & 0  & 0 & 1 \\
\hline
\end{tabular}
\caption{Truth table of the basic logic operations for a pair of
  inputs: $I_1, I_2$ \cite{mano}. Note that NAND and NOR are
  fundamental logic gates from which any logic circuit can be
  constructed \cite{mano}. Here AND and NAND, OR and NOR, XOR and XNOR
  are complementary pairs and are simply given by the NOT operation on
  the logic output, where NOT gives value $1$ when input is $0$
  and $0$ when input is $1$.\\}
\end{center}
\end{table}

\begin{table}
\begin{center}
\begin{tabular}{| c | c |  c |  c | c | c |}
 \hline Inputs & \ \ \ $S_{ext}$ \ \ \ & \ \ $A_{max}$ \ \ &NAND Logic with $A^*=1$&NOR Logic with $A^*=4$\\
\hline
$(0,0)$ & $0.0$ & $\approx 6.0$ & $1$ & $1$ \\ \hline
$(0,1)/(1,0)$ & $0.2$ & $\approx 2.0$ & $1$  & $0$ \\ \hline
$(1,1)$ & $0.4$ & $\approx 0.0$ & $0$ & $0$ \\
\hline
\end{tabular}
\caption{ The maximum concentration of protein $A$ mirroring the
  fundamental NAND and NOR logic gates. The output determination
  threshold $A^* = 1$ for the NAND logic operation and $A^* = 4$ for
  the NOR logic operation.\\}
\label{NAND_A_table}
\end{center}
\end{table}

\begin{table}
\begin{center}
\begin{tabular}{| c | c |  c | c | c |}
  \hline Inputs & \ \ \ $S_{ext}$ \ \ \ & \ \ \ $B_{max}$ \ \ \ &AND Logic with $B^*=30$&XNOR Logic with $B^*=15$\\
  \hline
  $(0,0)$ & $0.0$ & $\approx 20.0$ & $0$ & $1$ \\ \hline
  $(0,1)/(1,0)$ & $0.2$ & $\approx 4.0$ & $0$ & $0$ \\ \hline
  $(1,1)$ & $0.4$ & $\approx 45.0$ & $1$ & $1$ \\
  \hline
\end{tabular}
\caption{The maximum concentration of protein $B$ mirroring the
  AND and XNOR logic gates. The output determination threshold $B^* = 30$
  for the AND logic operation and $B^* = 15$ for the XNOR logic
  operation.\\}
\label{AND_B_table}
\end{center}
\end{table}

\begin{table}
\begin{center}
\begin{tabular}{| c | c |  c |  c |c | c |}
 \hline Inputs & \ \ \ $\kappa$ \ \ \ & \ \ \ $A_{max}$ \ \ \ &NAND Logic with $A^*=2$&NOR Logic with $A^*=6$\\
\hline
$(0,0)$ & $10.0$ & $\approx 9.0$ & $1$ & $1$ \\ \hline
$(0,1)/(1,0)$ & $30$ & $\approx 3.0$ & $1$ & $0$ \\ \hline
$(1,1)$ & $50$ & $\approx 0.0$ & $0$ & $0$ \\
\hline
\end{tabular}
\caption{The maximum concentration of protein $A$ mirroring the
  AND and XNOR logic gates. The output determination threshold $A^* = 2$
  for the NAND logic operation and $A^* = 6$ for the NOR logic
  operation.\\}
\label{kappa_A_table}
\end{center}
\end{table}

\begin{table}
\begin{center}
\begin{tabular}{| c | c |  c |  c | c | c |}
 \hline Input & \ \ \ $\kappa$ \ \ \ & \ \ \ $B_{max}$ \ \ \ &AND Logic with $B^*=40$&XNOR Logic with $B^*=20$\\
\hline
$(0,0)$ & $10.0$ & $\approx 28.0$ & $0$ &$1$ \\ \hline
$(0,1)/(1,0)$ & $30$ & $\approx 5.0$ & $0$ & $0$ \\ \hline
$(1,1)$ & $50$ & $\approx 45.0$ & $1$ & $1$ \\
\hline
\end{tabular}
\caption{The maximum concentration of protein $B$ mirroring the
  AND and XNOR logic gates. The output determination threshold $B^* = 40$
  for the AND logic operation and $B^* = 20$ for the XNOR logic
  operation.\\}
\label{kappa_B_table}
\end{center}
\end{table}

\bigskip
\bigskip
\bigskip

\newpage

\begin{thebibliography}{1}
\bibitem{computing} J. P. Crutchfield, W. L. Ditto, and S.  Sinha,
  Introduction to Focus Issue: Intrinsic and Designed Computation:
  Information Processing in Dynamical Systems -- Beyond the Digital
  Hegemony, {\em Chaos} {\bf 20} 037101 (2010)

\bibitem{chaogates} S. Sinha and W.L. Ditto, {\em Phys. Rev. Lett.}
  {\bf 81} (1998) 2156; K. Murali, S. Sinha, W.L. Ditto, A.R. Bulsara,
  {\em Phys. Rev. Lett.} {\bf 102} (2009) 104101; W. L. Ditto,
  A. Miliotis, K. Murali, S. Sinha, and M. L. Spano, {\em Chaos} {\bf
    20} (2010) 037107

\bibitem{quantum} A. Steane (1998) {\em Rep. Prog. Phys.} {\bf 61} 117

\bibitem{dna}{\em DNA Computing: New Computing Paradigms}, G.  Paun,
  G. Rozenberg and A. Salomaa, Springer (Berlin and New York) 1998

\bibitem{benenson} Y. Benenson, {\em Science} {\bf 340} (2013) 554 

\bibitem{ando} Ando, H., Sinha, S., Storni, S., and Aihara, K.:
  Synthetic gene networks as potential ﬂexible parallel logic
  gates. Europhys. Letts., {\bf 93} (2011) 50001.

\bibitem{gene2} E. H. Hellen, S. K. Dana, J. Kurths, E. Kehler,
  S. Sinha, Noise-aided Logic in an Electronic Analog of Synthetic
  Genetic Networks, {\em PLoS One} (2013)
  doi:10.1371/journal.pone.0076032

\bibitem{dockery}Dockery, J.D., Keener, J.P.:A mathematical model for
  quorum sensing in Pseudomonas aeruginosa. Bull. Math. Biol. 63,
  95–116 (2001).

\bibitem{james} S. James, P. Nilsson, G. James, S. Kjelleberg and
  T. Fagerström (2000). Luminescence control in the marine bacterium
  Vibrio ﬁscheri: An analysis of the dynamics of lux
  regulation. J. Mol. Biol. 296, 1127-1137.

\bibitem{ward} Ward, J.P., King, J.R., Koerber,A.J.,Williams, P.,
  Croft, J.M., Sockett, R.E.:Mathematical model of quorum sensing in
  bacteria. IMA J. Math. Appl. Med. Biol. 18, 263–292 (2001).

\bibitem{Hellen} Hellen EH, Dana SK, Zhurov B, Volkov E (2013)
  Electronic Implementation of a Repressilator with Quorum Sensing
  Feedback. PLoS ONE 8(5): e62997.  doi:10.1371/journal.pone.0062997.

\bibitem{strogatz} Garcia-Ojalvo, J., Elowitz, M.B. and Strogatz,
  S.H.: Modeling a synthetic multicellular clock: Repressilators
  coupled by quorum sensing. Proceedings of the National Academy of
  Sciences of the U.S.A. 101, 10955-10960 (2004).

\bibitem{potapov} Potapov, I., Zhurov, B. and Volkov, E.: “Quorum
  sensing” generated multistability and chaos in a synthetic genetic
  oscillator.  Chaos,22, 023117 (2012).

\bibitem{koseska} Koseska A, Ullner E, Volkov E, Kurths J, Garcı
  ´a-Ojalvo J (2010) Cooperative differentiation through clustering in
  multicellular populations. J. Theor. Biol.  263: 189–202.

\bibitem{Elowitz} Elowitz M, Lim WA (2010) Build life to understand
  it. Nature 468: 889–890.

\bibitem{mano} M.M. Mano, {\em Computer System Architecture}, 3rd
  edition, Prentice Hall, Englewood Cliffs, 1993; Bartee, T.C. {\em
    Computer Architecture and Logic Design}, New York, Mc-Graw Hill,
  1991.
  
\end{thebibliography}
\end{document}